\documentclass[english,ngerman]{bvm} 

\bvmdef\articlenumber{3251}

\bvmdef\type{V}

\date{}

\title{Does Proprietary Software Still Offer Protection of Intellectual Property in the Age of Machine Learning?}

\titlerunning{Offer Proprietary Algorithms Still Protection of Intellectual Property?}

\subtitle{A Case Study using Dual Energy CT Data}

\author{Andreas Maier$^1$, Seung Hee Yang$^2$, Farhad Maleki$^3$, Nikesh Muthukrishnan$^3$, Reza Forghani$^3$}

\authorrunning{Maier et al.} 

\institute{%
$^1$Pattern Recognition Lab, FAU Erlangen-N\"urnberg\\
$^2$Department Artificial Intelligence in Medical Engineering, FAU Erlangen-N\"urnberg\\
$^3$McGill University Hospital, McGill University
}

\email{
andreas.maier@fau.de
}

\begin{document}

%
\selectlanguage{english}

\maketitle

\begin{abstract}
In the domain of medical image processing, medical device manufacturers protect their intellectual property in many cases by shipping only compiled software, i.e. binary code which can be executed but is difficult to be understood by a potential attacker. In this paper, we investigate how well this procedure is able to protect image processing algorithms. In particular, we investigate whether the computation of mono-energetic images and iodine maps from dual energy CT data can be reverse-engineered by machine learning methods. Our results indicate that both can be approximated using only one single slice image as training data at a very high accuracy with structural similarity greater than 0.98 in all investigated cases.
\end{abstract}

\section{Introduction}
Medical image processing and analysis is heavily relying on device manufacturers as any treatment to patients has to follow medical device legislation \cite{sorenson2014improving}. As such vendors have great interest to protect their intellectual property such that it cannot be easily reproduced by their competition. This often results in monolithic software packages which are shipped in binary software only such that analyses can be performed while the underlying algorithm is not accessible to third parties.

As reverse engineering attacks are very expensive and require in depth knowledge \cite{canfora2011achievements}, image processing algorithms are generally assumed to be robust against these attacks by companies in the healthcare sector. Yet, in the age of machine and deep learning \cite{lecun2015deep,maier2019gentle}, this may no longer hold since very complex processing systems can be approximated by such techniques.

In this paper, we investigate how well image processing algorithms can be approximated using standard open source software. In particular, we will use the learning module provided in CONRAD \cite{maier2013conrad}, which is based on the WEKA machine learning toolbox \cite{hall2009weka}, to learn algorithms related to dual energy imaging. In the first attack case, we look into computation of mono-energetic images from dual energy data given the same sampling grid. In the second case, we look into non-linear material decomposition estimated from manually pre-processed images as commonly found in clinical practice.

\section{Methods}
Dual energy CT is based on the acquisition of X-ray images acquired at different kilovoltage peaks (kVp) \cite{maier2018medical}. Due to the differences in physical absorption, the materials in the imaged volume can be analysed in a quantitative way based on their physical spectral properties allowing advantages with respect to medical image analysis \cite{chen2020automatic}. In particular, reconstruction of virtual monochromatic images and iodine maps are very common in this framework. The acquisition using already two photon energies ($I_\text{low}$ and $I_\text{high}$) allows to estimate a wide spectrum of other acquisition energies virtually $I_\text{virt}$ as described in \cite{krauss2011dual}:
\begin{equation}
I_\text{virt} = \alpha \cdot I_\text{high} + (1-\alpha) I_\text{low}
\label{eq1}
\end{equation}
Above approximation is quite common in the world of medical physics and offers the straight forward hypothesis that also commercial algorithms should be explainable using a linear model.

Dual energy imaging also allows to perform material decomposition such as the extraction of water and iodine maps \cite{liu2009quantitative}. The relation between the actual energies and the materials is generally non-linear and is typically estimated using polynomial estimators. Several authors \cite{lu2015projection,lu2019learning,geng2020pms} suggested that a general machine learning model $f$ is also suited to perform the decomposition into material maps $I_\text{mat}$:
\begin{equation}
I_\text{mat} = f(I_\text{high}, I_\text{low})
\label{eq2}
\end{equation}
As such, we are to expect general non-linear computations in material decomposition algorithms.

\section{Experiments and Results}

\begin{figure}[tbp]
\includegraphics[width=\linewidth]{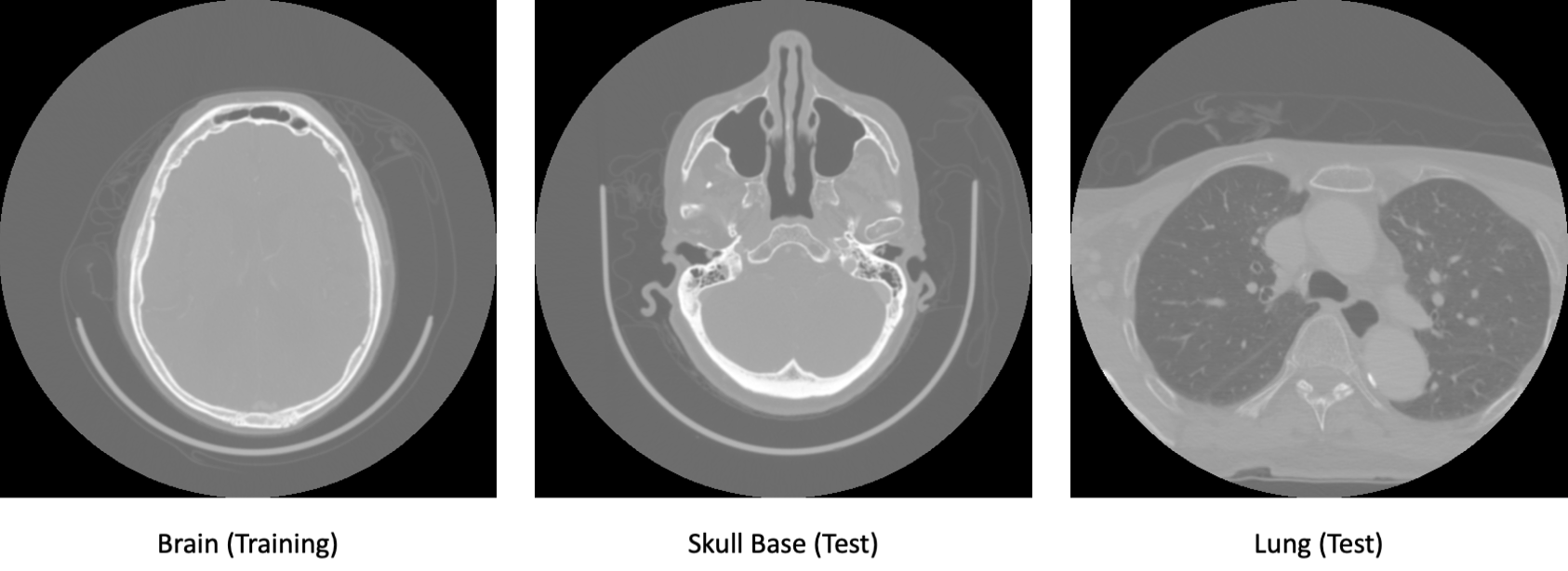}
\caption{All machine learning models were only estimated using a single slice through the brain in this paper. Evaluations were performed on a section through the skull base and the lung.} 
\label{fig-1}
\end{figure}

In order to investigate above dual energy algorithms, we explored the dual energy data from a clinical scan of the neck (which includes part of the skull base and brain superiorly and part of the lungs inferiorly) using machine learning algorithms. For the training, one axial slice of the brain was selected. For testing one axial slice through the skull base and one axial slice through the lung were chosen in order to demonstrate that the learned parameters generalize over different anatomical regions as displayed in Fig.~\ref{fig-1}. The acquisition protocol used fast kVp-switching between 140 and 80 kV. 

The original raw DICOM data only contained one of the acquisition energies in the respective DICOM field. Yet, the DICOM header offered four additional proprietary DICOM fields which stored additional pixel data in 16-bit encoding which was extracted using a custom DICOM reader supplied within the CONRAD software framework \cite{maier2013conrad}. Amongst the extracted data, we found the second energy image as well as additional correction factors which are apparently used by the vendor to compute the resulting mono-energetic and material decomposition images.

All machine learning methods were trained with pixel-wise correspondences, as one would expect for single energy computation or material decomposition. None of the approaches considered a local neighborhood which implies that changes in resolution and operations such as noise reduction cannot be modelled by the learning approach. All experiments used the default parameter setting of WEKA. No additional parameter tuning was applied.

\subsection{Single Energy Estimation}

In order to explore how well the computation of mono-energetic images is protected, single energy slices from 40 to 140 keV in steps of 20 keV were created. As observed in Eq.~\ref{eq1}, we assume a linear relation between the dual energy raw data and the produced results. 

The results summarized in Table~\ref{tab1} indicate that the linear model is able to recover all parameters of the mono-energetic image computation correctly. Correlation coefficient $r$ and structural similarity (SSIM) lie above $0.999$ in all cases indicating that the computation is identical up to 16-bit precision.

\begin{table}[t]
\caption{All mono-energetic slices could be reproduced with correlations and SSIMs greater than $0.999$ which essentially shows that the ground truth algorithm could be extracted.}
\label{tab1}
\begin{tabular*}{\textwidth}{l@{\extracolsep\fill}ccc}
\hline
Mono-energetic Image Estimation                & $r$ & ~~SSIM~~   & ~~~Slice~~~ \\
\hline
40 kV            & 0.999       & 0.999  & Skull    \\
40 kV            & 0.999       & 0.999  & Lung   \\
60 kV            & 0.999       & 0.999  & Skull    \\
60 kV            & 0.999       & 0.999  & Lung   \\
80 kV            & 0.999       & 0.999  & Skull    \\
80 kV            & 0.999       & 0.999  & Lung   \\
100 kV           & 0.999       & 0.999  & Skull    \\
100 kV           & 0.999       & 0.999 & Lung   \\
120 kV           & 0.999       & 0.999  & Skull    \\
120 kV           & 0.999       & 0.999  & Lung   \\
140 kV           & 0.999       & 0.999  & Skull    \\
140 kV           & 0.999       & 0.999  & Lung   \\
\hline
\end{tabular*}
\end{table}

\begin{figure}[tbp]
\includegraphics[width=\linewidth]{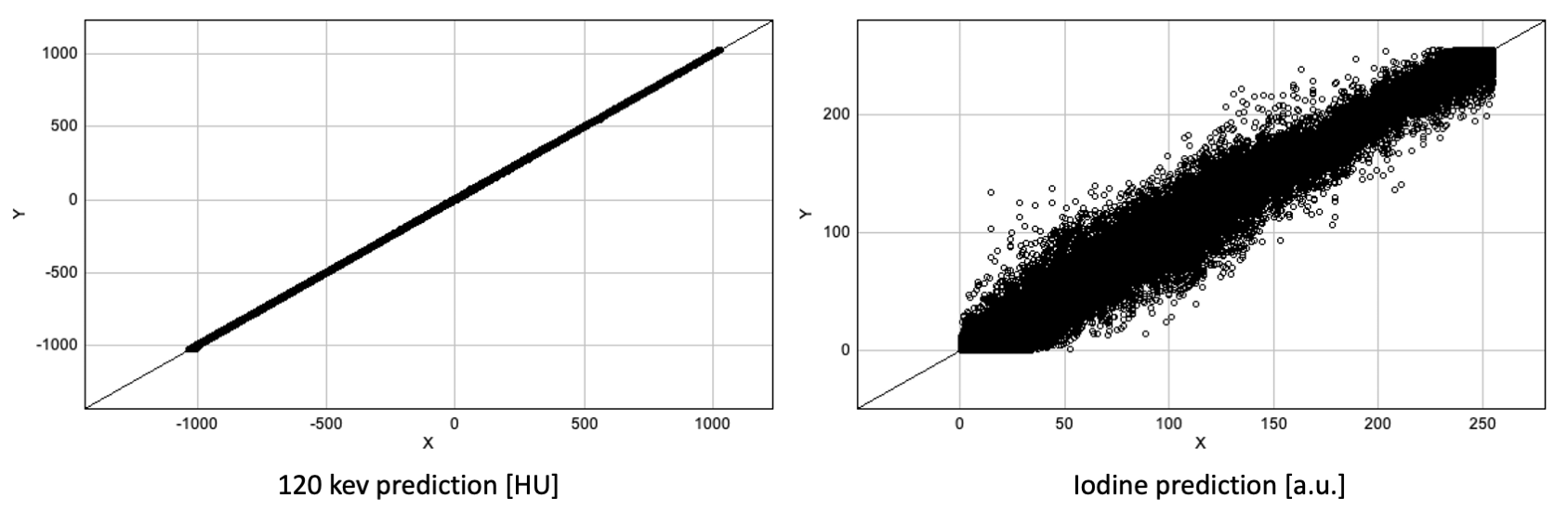}
\caption{Regression analysis on the lung slice data set: The left image shows the strong correlation between ground truth values and the machine learning estimates for the prediction of 120 kV images. The right side shows a scatter analysis between the ground truth iodine maps and the predictions by the tree model with an SSIM of more than $0.98$.} 
\label{fig-2}
\end{figure}

\subsection{Iodine Map Prediction}

For the reliability analysis of the material decomposition, we created an Iodine map using the software on the clinical work station. In order the make the attack task even more realistic, the image resolution was and dimensions were changed from $512 \times 512$ to $1200 \times 1024$. To estimate the correct slice orientation, images were manually registered and re-sampled to match the raw data orientation. Based on the alignment, a linear and a non-linear machine learning estimator was trained. For the non-linear model WEKA's Reduced Error Pruning (REP) Tree \cite{hall2009weka} was chosen as it is faster than random forests while being more robust than the classical tree learners.

The results shown in Table~\ref{tab2} indicate that a linear model is not able to describe the material decomposition entirely yielding correlation and SSIM values of only $0.88$ even in the case of the training slice in the brain area. The non-linear model performs much better with consistent correlation and SSIM values of over $0.98$. Visual analysis of the predicted Iodine image and the ground truth indicated that the mismatch mainly comprised in changes of noise patterns, image resolution, and image alignment. Hence, the REP Tree model delivers a very close approximation of the ground truth algorithm of the vendor.

\begin{table}[t]
\caption{The linear model only creates a coarse estimation of the material decomposition. The non-linear REP Tree, however, consistently achieves correlation and SSIM values greater than §0.98§. The main difference between the produced images in this case is the change of orientation and resolution.}
\label{tab2}
\begin{tabular*}{\textwidth}{l@{\extracolsep\fill}ccc}
\hline
Mono-energetic Image Estimation                & $r$ & ~~SSIM~~   & ~~~Slice~~~ \\
\hline
Iodine (linear)  & 0.894       & 0.888  & Brain     \\
Iodine (linear)  & 0.825       & 0.808  & Skull    \\
Iodine (linear)  & 0.766       & 0.475  & Lung   \\
\hline
Iodine (REP Tree) & 0.997       & 0.997  & Brain     \\
Iodine (REP Tree) & 0.993       & 0.993  & Skull    \\
Iodine (REP Tree) & 0.985       & 0.985  & Lung  \\
\hline
\end{tabular*}
\end{table}

\section{Discussion}

The presented results indicate that linear prediction algorithms can be approximated by pixel-wise machine learning methods very well. Furthermore, even in adverse conditions in a non-linear setting with changes in resolution and slice alignment, very high correlation values and structural similarity could be obtained using a tree learning. We assume that advanced machine learning models \cite{maier2019gentle} would be able to capture such differences as well. Yet, they would probably require more than a single slice image as training data.

Generally, it is surprising to see how common open source tools already are able to approximate vendor algorithms. As such one might wonder whether closed source software still offers sufficient protection of algorithms and their intellectual property. In the machine learning world, even industrial players started to publish source code and trained models. One reason for this is of course that doing so allows them to incorporate contributors outside their payroll. Open-sourcing such software with appropriate licenses that protect the vendor's intellectual property might be a more promising approach in comparison to closed-source software that can be easily reverse-engineered.

Given the results in this paper, the advantage of closed source in terms of algorithm protection might not be as strong as one might expect. Hence, even industrial players in the field of medical image processing may want to reconsider the disadvantages of open source with respect to software and intellectual security. Given enough data, we expect many other algorithms could fall to the presented attack scheme.

Developing in open source domain may be even of advantage in the future for many medical device manufacturers as they would be able to harness the ingenuity of the entire academic community. The authors of this paper strongly believe that this direction would be beneficial for the entire medical image processing community. 

\section{Conclusion}

Closed source software is no longer a sufficient protection strategy for many image processing algorithms as demonstrated by the results in this paper. Even non-linear material decomposition algorithms can be estimated by rather simple means in adverse conditions using only a single slice image. Generally, results yielded correlation and structural similarity of $0.98$ or higher which indicates that the machine learning methods were able to precisely estimate the closed source algorithms correctly.

\bibliographystyle{bvm}

\bibliography{3251}

\marginpar{\color{white}E\articlenumber}

\end{document}